\documentclass[12pt]{article}

\usepackage{natbib}

\usepackage{amsmath}
\usepackage{amssymb}
\usepackage{graphicx}
\usepackage{authblk}
\usepackage{xcolor}
\usepackage{setspace}
\usepackage[margin = 1in]{geometry}
\usepackage{caption}
\usepackage{subcaption}
\usepackage{booktabs}
\usepackage{siunitx}
\usepackage{url}

\usepackage{lineno}

\newcommand{\bs}{\boldsymbol{s}}
\newcommand{\ds}{\delta_s}
\newcommand{\dt}{\delta_t}

\newcommand{\bw}{\boldsymbol{w}}
\newcommand{\by}{\boldsymbol{y}}
\newcommand{\bz}{\boldsymbol{z}}
\newcommand{\boldeta}{\boldsymbol{\eta}}

\newcommand{\bSigma}{\boldsymbol{\Sigma}}
\newcommand{\bQ}{\boldsymbol{Q}}
\newcommand{\btheta}{\boldsymbol{\theta}}
\newcommand{\bB}{\boldsymbol{B}}
\newcommand{\bm}{\boldsymbol{m}}
\newcommand{\bD}{\boldsymbol{D}}
\newcommand{\bone}{\boldsymbol{1}}
\newcommand{\Id}{\boldsymbol{I}}
\newcommand{\bomega}{\boldsymbol{\omega}}
\newcommand{\bOmega}{\boldsymbol{\Omega}}

\newcommand{\E}{\mathrm{E}}
\newcommand{\Cov}{\mathrm{Cov}}
\newcommand{\Var}{\mathrm{Var}}

\newcommand{\simiid}{\overset{\mathrm{iid}}{\sim}}

\newcommand{\calS}{\mathcal{S}}
\newcommand{\calT}{\mathcal{T}}
\newcommand{\calO}{\mathcal{O}}
\newcommand{\calP}{\mathcal{P}}

\newcommand{\bzero}{\boldsymbol{0}}

\title{Spatial-temporal prediction of forest attributes using latent Gaussian models and inventory data}

\author[1*]{Paul B. May}
\author[2]{Andrew O. Finley}

\affil[1*]{South Dakota School of Mines \& Technology\protect\\
Department of Mathematics\protect\\
Rapid City, South Dakota, United States \protect\\
paul.may@sdsmt.edu \vspace{10pt}}
\affil[2]{Michigan State University\protect\\
Department of Forestry -- Department of Statistics and Probability \protect\\
East Lansing, Michigan, United States\protect\\
finleya@msu.edu}
\date{}

\begin{document}

\maketitle


\begin{abstract}
    The USDA Forest Inventory and Analysis (FIA) program conducts a national forest inventory for the United States through a network of permanent field plots. FIA produces estimates of area averages and totals for plot-measured forest variables through design-based inference, assuming a fixed population and a probability sample of field plot locations. The fixed-population assumption and characteristics of the FIA sampling scheme make it difficult to estimate change in forest variables over time using design-based inference. We propose spatial-temporal models based on Gaussian processes as a flexible tool for forest inventory data, capable of inferring forest variables and change thereof over arbitrary spatial and temporal domains. It is shown to be beneficial for the covariance function governing the latent Gaussian process to account for variation at multiple scales, separating spatially local variation from ecosystem-scale variation. We demonstrate a model for forest biomass density, inferring 20 years of biomass change within two US National Forests.
\end{abstract}


\section{Introduction}

The Forest Inventory and Analysis (FIA) program of the USDA Forest Service monitors forest resources across the United States through a network of spatially random, permanent field plots. Measurements at these field plots provide an invaluable data source for estimating key forest attributes such as biomass, carbon stocks, and species composition. FIA produces estimates of area averages and totals through design-based estimation methods \citep{bechtold2005enhanced}. Unlike model-based approaches, which may rely on assumed distributions, relationships to auxiliary data and spatial-temporal behavior, design-based methods depend only on the sampling design \citep{gregoire1998design}. Because the FIA plot network was established under a predetermined sampling design (though deviations may occur due to logistics and denial of access to private land), the corresponding design-based estimators are ensured robust inference when sample size is sufficiently large.

The current FIA design-based protocols have some limitations. First, the precision of area estimates is dependent on the sample size within the area. FIA stakeholders in both the public and private sector are often in need of estimates at finer scales than the program's sampling density allows \citep{wiener2021united, prisley2021needs}. This need has sparked a rapid growth in Small Area Estimation (SAE) techniques for forest inventory data in the scientific literature. Most SAE techniques are model-based, using assumed relationships to auxiliary data and spatial correlation to ``borrow strength'' from a broader training domain, increasing the effective sample size in the target estimation area \citep{ver2018hierarchical, temesgen2021using, white2021hierarchical, may2023spatially, white2024small, shannon2024toward}.

A second limitation of FIA design-based estimates is more directly addressed in this work. For many applications, what is needed is an estimate of forest attributes at a particular point in time, or change across a time span \citep{prisley2021needs}. FIA collects measurements in a periodic inventory, systematically remeasuring all plots across a multi-year cycle. A full cycle of measurements completes the random sample from which design-based estimates can be constructed. These estimates assume the population is fixed, but in reality the population change over the course of the cycle may be physically substantial. One may intuit the estimate to represent the average population over the time period of the cycle, but over a rapidly changing area, different measurement timetables could produce quite different estimates, a source of uncertainty not considered in the design-based estimates.

Alternatively, one could consider constructing annual estimates using only measurements within a particular year. The severely truncated sample size aside, due to the systematic re-measurement timetable of the plots, the annual samples poorly approximate a simple random sample and the annual sample sets within a cycle are strongly dependent. This latter issue is especially problematic, as it can create cyclical patterns in the annual estimates, making even informal assessments of inter-cycle change difficult. See \cite{moisen2020estimating}, especially Section 3.2.1, for a more detailed discussion of these challenges. Model-based, areal methods have been proposed, using the design-based, annual estimates as a model response and fixed effects and temporal random effects to more precisely infer population averages \citep{hou2021updating, shannon2024toward}. But the focus of these studies has been increasing the effective sample size within each year, and they do not address the difficult behavior of annual estimates discussed above.\par
We propose latent Gaussian models (LGMs), using individual plot measurements as the model response, as an alternative framework for FIA statistical analyses. By modeling at the plot level, we avoid reliance on design-based area estimates, providing a self-contained approach that is less constrained by the sampling design. We use Gaussian processes, continuously indexed over space and time, as our Gaussian random effects. LGMs with Gaussian processes are widely popular in geostatistics \citep{gelfand2016spatial, opitz2017latent}, and have seen application in forest inventory data for spatial-only analyses, assuming temporal stasis over the measurement period \citep{finley2011hierarchical, hunka2025geostatistical}. Extending this to spatial-temporal analyses requires covariance structures that can skillfully capture relationships across both space and time, which may be complex. We show in this work that for forest inventory data it is beneficial for the assumed structure to account for variation at multiple scales, separating local variation from broader, ecosystem-scale variation. Finally, we take a Bayesian approach to inference, computing posterior probability distributions representing beliefs on target quantities given the observed data. These posterior distributions can be flexibly summarized into interpretable probability statements useful to land managers and policy makers.\par
We demonstrate an LGM for above-ground biomass density (AGBD, mass of the trees above ground divided by occupied land area) across two separate ecoregions in the United States. This model is used to infer AGBD trends within two federally owned National Forests within the ecoregions. We provide validation of the model and justification of proposed space-time covariance structure through a cross-validation study. The primary goal of this demonstration is to provide an example of rigorous inference on time-specific forest population values and change in population values, but we also provide discussion on how covariates can be included into the model to improve inference on status and change at finer spatial-temporal scales.



\section{Data}\label{sec:data}

\subsection{The Forest Inventory and Analysis program}\label{sec:fia}

\begin{figure}[htb]
    \centering
    \includegraphics[width=\linewidth]{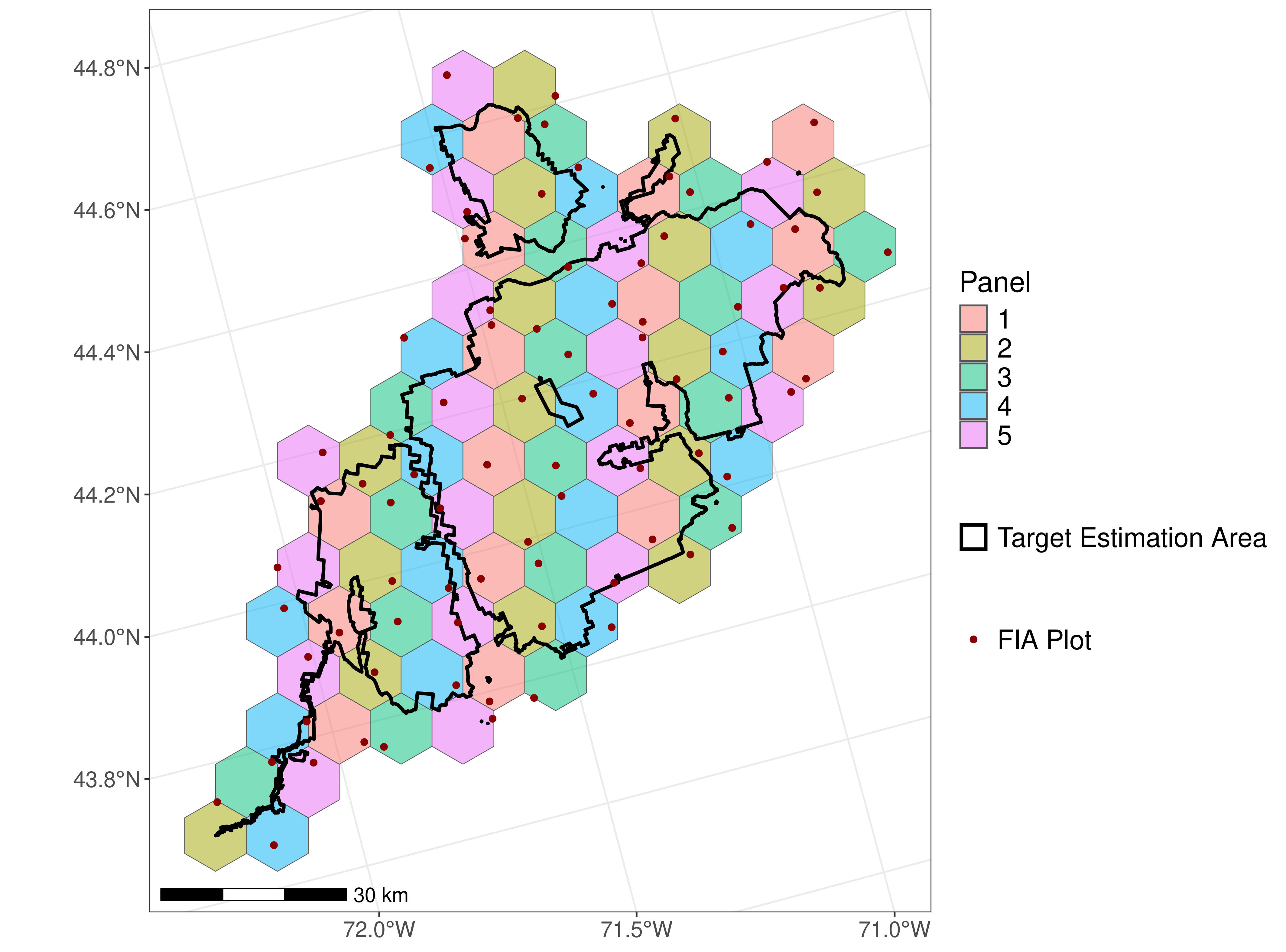}
    \caption{An illustration of the FIA sampling scheme with a 5-year cycle over White Mountain National Forest. A permanent plot location was randomly selected within each 6,000 acre ($\approx\text{24 km}^2$) hexagon. Each year, all plots corresponding to a single panel are measured. Over a full cycle (e.g., 2001--2004), all plots will have been measured, yielding an approximately simple random sample over space. Measurements within a target estimation area are used to produce a design-based estimate for that area and cycle.}
    \label{fig:fiasampling}
\end{figure}

FIA maintains over 300,000 permanent field plots across the contiguous United States, on which a suite of important forest attributes are periodically measured. The plot locations were selected according to a multistage sampling design, which is assumed to well-approximate a simple random sample (Figure \ref{fig:fiasampling}). In the first stage, a regular grid of 6,000 acre hexagons was randomly placed across the landmass. Within each hexagon, a single plot location was uniform-randomly selected. To create a timetable for plot measurements, the hexagons were systematically assigned to a finite number of panels, with 5 or 7 panels for the Eastern US and 10 panels for the Western US. Each year, all plots corresponding to a single panel are measured. After a complete cycle (5 or 7 years in the East, 10 years in the West), all plots have been measured, completing the simple random sample. A new cycle then begins, remeasuring each plot on a deterministic timeline. FIA sometimes commits high-priority areas to increased spatial sampling intensities. For a $M$-fold intensification, every hexagon in the intensified area is partitioned into $M$ subareas. The $M-1$ subareas of each hexagon that do not already contain a plot have a plot uniform-randomly added. These added plots inherit the panel and measurement timeline of the containing hexagon.

A primary objective of FIA is to estimate averages and totals of forest attributes over target areas. FIA does so through design-based methods, where inference is conducted through a hypothetical frequentist resampling of the plot locations over a fixed population. When using a complete cycle of measurements, the fixed-population assumption may poorly represent reality, especially considering the 10 year cycle for the West. Using only a single year of data provides more time-specificity, but severely truncates the sample size, and the resulting sample does not well-approximate a simple random sample due to the systematic arrangement of the panels. For similar reasons, assessing inter-cycle population change through consecutive annual estimates is made difficult by dependencies between the panel samples (e.g., Panel 1 locations inform possible Panel 2--5 locations).

\subsection{Study area}\label{sec:studyarea}

\begin{figure}[tb]
    \centering
    \includegraphics[width=\linewidth]{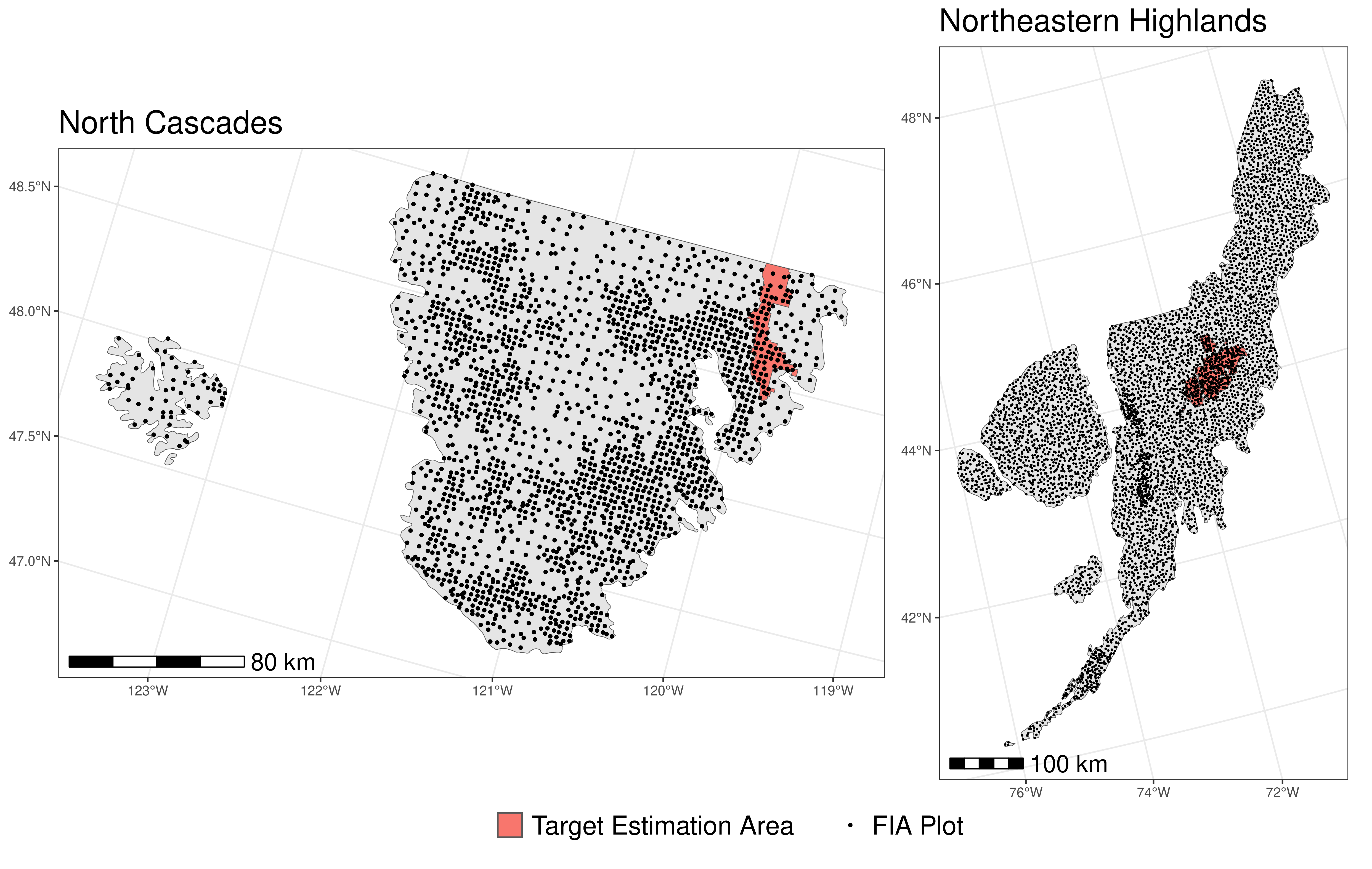}
    \caption{The two ecoregions selected to demonstrate the methods (Cascades in the Pacific northwest and Highlands in the northeast), each with a target estimation area of particular interest (Colville National Forest and White Mountain National Forest respectively).}
    \label{fig:studyarea}
\end{figure}

We use two EPA Level 3 ecoregions (\citealp{omernik1987ecoregions}, \url{https://www.epa.gov/eco-research/ecoregions}) as our demonstration areas, using all available FIA plot measurements from 2001 to 2021 (Figure \ref{fig:studyarea}). The North Cascades is an ecoregion in the Western US, with the vast majority of the land owned and managed by the federal government. The region contains 2,150 unique plot locations with scattered areas of increased sampling intensity, giving an average of one plot per 14 km\textsuperscript{2}. Over the considered timeframe, there have been 3,162 plot measurements, with 71\% of the plots measured more than once, and no plots measured more than twice due to the 10 year cycle. Mostly forested, 84\% of the plot measurements reported non-zero forest AGBD, with 6\% of the plots possessing more than one measurement transitioning from non-zero AGBD to zero AGBD or vice versa. Out of the forested plot measurements, the average measured AGBD is 170 Mg/ha with a maximum measurement of 1,360 Mg/ha (only 5 plots in the US have a measurement larger than this, all of which are on the Northern California coast). Our target estimation area is the section of Colville National Forest within the North Cascades (Colville has another disjoint section east of the North Cascades ecoregion).\par 
The second analyzed region is the Northeastern Highlands in the Eastern US, where the majority of the land is privately owned. The region contains 5,631 unique plot locations with little intensified sampling, giving an average of one plot per 22 km\textsuperscript{2}. Over the considered timeframe, there have been 18,846 measurements with a median of 3 measurements per plot and a maximum of 5. Also mostly forested, 90\% of measurements reported non-zero forest AGBD, with 4\% of the plots possessing more than one measurement transitioning from non-zero AGBD to zero AGBD or vice versa. Out of the forested plot measurements, the average measured AGBD is 109 Mg/ha with a maximum measurement of 497 Mg/ha. Our target estimation area for this ecoregion is White Mountain National Forest.

\section{Methods}\label{sec:methods}

For design-based inference, the population throughout a target estimation area, $y(\bs),~\forall\bs\in\mathcal{A}$, where $\bs$ is a two-dimensional spatial coordinate vector, is considered fixed and non-stochastic, and the sampled locations assumed to follow some sampling design $\bs_1,\ldots,\bs_n\sim P_\text{design}$. In contrast, for a spatial-temporal, model-based approach, the measurement locations and times are assumed fixed and the population considered random, $y(\bs,t)\sim P_\text{model},~\forall (\bs,t)\in \calS\times\calT$. Note that in this case it may be useful to consider some greater spatial domain than the target estimation area, $\mathcal{A} \subset \calS$, and greater timeframe, if we are willing to assume the greater domain follows a common model.\par 
For design-based inference, the assumed probability law is motivated by the execution of the sampling. In fact, this is often best understood conversely, in that the sampling was executed according to a predetermined probability law. For model-based inference, the probability law is more `invented' according to our beliefs about the data and the behavior of the generating physical process, possibly over the course of multiple iterations of model development and validation. As much as possible, the assumed model should capture aspects of the process that will be pertinent to prediction or other analysis goals. In particular to forest inventory data, the model should capture marginal properties of the data (many attributes are zero-inflated due to non-forested plots, AGBD is positively skewed, tree species data can be compositional or multinomial), as well as multivariate properties, depicting relationships across space and time.   

\subsection{Spatial-temporal Gaussian processes and LGMs}\label{sec:stgp}

Gaussian processes are central to geostatistics because of their ability to capture spatial-temporal correlations in an analytically tractable form. Process $w(\bs, t)$ is a Gaussian process on $\calS \times \calT$ if for any finite set of points in space-time, the corresponding vector $[w(\bs_1, t_1) \cdots w(\bs_n, t_n)]$ is multivariate normal. Such allows modeling covariances between our observation coordinates, and the closure of the normal distribution upon conditioning facilitates prediction at arbitrary coordinates. Because normality assumptions are untenable for many data, the Gaussian process is often embedded in a data model, $y(\bs,t) \simiid f\left(y |w(\bs,t)\right)$, forming a LGM in totality. The latent Gaussian process provides an underlying link across space and time, allowing informative predictions of $y(\bs,t)$ at unobserved locations and times, area averages at specific times, and change across time. \par
A Gaussian process is fully defined by a mean and covariance function. We define $\E[w(\bs,t)]=0$, allowing other model components to dictate a systematic mean. A covariance function takes any two coordinates and returns the covariance between them. To be valid, the covariance function must be positive definite, meaning for any set of $n$ coordinates the resulting $n\times n$ covariance matrix is positive definite \citep{genton2001classes}. In this work, we use stationary and isotropic covariance functions, meaning the function depends only on the distance in space and time
\begin{equation}
    K(\delta_s, \delta_t) \equiv \Cov\left[w(\bs, t), w(\bs', t')\right],\quad \forall (\bs, t), (\bs', t') \in \calS \times \calT,
\end{equation}
where $\ds = \|\bs - \bs'\|$ and $\dt = |t - t'|$ are the spatial and temporal lag respectively. For most applications lacking periodic behavior, the assumed covariance functions are monotone decreasing with $\ds$ and $\dt$ and bounded below by zero: observations nearer in space and time are expected to be more similar. Devising valid and realistic spatial-temporal covariance functions is difficult and much research has been dedicated to the subject \citep{cressie1999nonseparable, gneiting2002nonseparable, porcu202130}. The simplest method is to construct `separable' functions,
\begin{equation}\label{eq:separblecov}
    K(\ds,\dt) = \sigma^2 K_\calS(\ds)\cdot K_\calT(\dt) ,
\end{equation}
where $\sigma$ is a standard deviation parameter and $K_\calS(\ds)$ and $K_\calT(\dt)$ are valid covariance functions on $\calS$ and $\calT$ respectively, normalized into correlation functions so that $K_\calS(0)=K_\calT(0) = 1$. The Mat\'ern class is popular for both spatial and temporal correlation functions. For fixed smoothness $\nu=1/2$, the Mat\'ern class reduces to the exponential class
\begin{equation}\label{eq:exponentialcor}
    K_\calS(\ds) = \exp\left(-\frac{\ds}{\phi}\right),\quad K_\calT(\dt) = \exp\left(-\frac{\dt}{\lambda}\right),
\end{equation}
where $\phi,\lambda>0$ are range parameters dictating the rate of exponential decay in space and time. Regardless of the individual correlation functions selected, separable space-time functions have the drawback of lacking space and time interactions: for any two fixed temporal lags, $d_t$ and $d_t'$, the curves $K(\ds, d_t)$ and $K(\ds, d_t')$ differ only by a multiplicative constant, and vice versa for fixed space lags.\par
The semivariogram is a function related to the covariance function,
\begin{equation}
    \gamma(\ds, \dt) \equiv \frac{1}{2}\Var[w(\bs, t) - w(\bs', t')] = K(0,0) - K(\ds,\dt).
\end{equation}
The advantage of the semivariogram is we can compute a model-free, empirical semivariogram with observed data, $y(\bs_1,t_1),\ldots,y(\bs_n,t_n)$. A space of considered spatial-temporal lags are partitioned into a set of rectangular bins, $B_1,\ldots,B_K\subset \mathbb{R}^+\times\mathbb{R}^+$. Then, the sample variance of pairwise-differences are computed within each bin,
\begin{equation}
    \hat{\gamma}(B_k) = \frac{1}{2|B_k|}\sum_{(\ds^{(ij)}, \dt^{(ij)})\in B_k} \left(y(\bs_i,t_i) - y(\bs_j,t_j)\right)^2. \label{eq:empvario}
\end{equation}
The empirical semivariogram can only be computed with the observed $y(\bs,t)$ and not the latent $w(\bs,t)$, and will therefore contain any unstructured, random variance from $f\left(y|w(\bs,t)\right)$, sometimes referred to as the ``nugget.'' While the direct use of the semivariogram for inference has dwindled in favor of likelihood-based techniques \citep{stein1999interpolation}, it remains a useful tool for exploratory analysis and we will use it to demonstrate interesting spatial-temporal properties of forestry data.\par
Finally, a difficulty in modeling with Gaussian processes is the computational demand, requiring the solution of $n\times n$ linear systems and log-determinants, where $n$ is the number of observations. We use Nearest Neighbor Gaussian Process (NNGP; \citealp{datta2016hierarchical}) representations of our Gaussian processes with $m=25$ neighbors (see Appendix \ref{app:nngp} for more details on our implementation), enabling fast inference through sparse matrix routines \citep{finley2019efficient}. Besides the NNGP, many other computationally efficient techniques exist that can yield high-fidelity representations of a native Gaussian process \citep{heaton2019case}.

\subsection{Proposed model}\label{sec:propmod}

Let $b(\bs, t)\geq 0$ represent forest AGBD at location $\bs$ and time $t$. The distribution of AGBD (and many other forest attributes) is semi-continuous, with substantial probability mass at zero due to non-forested areas, and positive, continuous support otherwise. Define the binary process $z(\bs, t)$ where $z(\bs,t) = 1 \iff b(\bs, t) > 0$. We assume the LGM
\begin{align}
    z(\bs,t) &\simiid \mathrm{Bernoulli}\left(\mu_z(\bs,t)\right), \\
    \mathrm{logit}\left(\mu_z(\bs,t)\right) &= \alpha_z + w_z(\bs, t),
\end{align}
giving a Bernoulli data model, for which the latent probability $\mu_z(\bs,t)$ is determined by a constant intercept $\alpha_z$ and a mean-zero Gaussian process $w_z(\bs, t)$. To model the continuous, positive attribute values, we first define $y(\bs,t) = g\left[ b(\bs,t)\right]$, where $g(\cdot)$ is a suitable transformation to account for the heavy positive-skew in the non-zero $b(\bs,t)$ values and yield a plausibly normally distributed $y(\bs,t)$:
\begin{align}
    y(\bs,t) &\simiid \mathrm{Normal}\left(\mu_y(\bs,t),~\tau^2\right),\\
    \mu_y(\bs,t) &= \alpha_y + w_y(\bs,t). \label{eq:ymu}
\end{align}
Here our data model is normal with standard deviation $\tau$ and mean determined by another intercept and Gaussian process. A final equation for AGBD can be written as $b(\bs,t) = g\left[y(\bs,t)\right]^{-1}z(\bs,t)$. The model can be thought of operating in two stages, the first predicting the probability of forest presence, and the second predicting, given forest presence, the magnitude of the forest attribute. A similar two-stage model was applied by \cite{finley2011hierarchical} and \cite{may2025calibrating} for continuous spatial modeling of forest inventory data and \cite{white2024small} for area-level modeling. Because forest presence over $\calS\times\calT$ will always be uncertain, the first binary stage will always be necessary. However, the second stage may be modified based on the target attribute, e.g., using a Poisson data model for tree count data.\par 
Ideally, the transformation $g[b(\bs,t)]=y(\bs,t)$ would be an invertible function with $g:\mathbb{R}^+\rightarrow\mathbb{R}$, making the logarithm a natural choice. However, we found the logarithm to be too strong, resulting in a negatively-skewed $y(\bs,t)$. We prefer root functions, $g(b)=b^{1/r};~r\in\mathbb{N}$, which can be tailored to the ecoregion, with stronger roots for high biomass regions and weaker roots for moderate biomass regions. The demerit of this choice is that for odd roots the inverse function maps negative $y(\bs,t)$ values to negative $b(\bs,t)$ values and for even roots the inverse function is not one-to-one on $\mathbb{R}$. However, we have found that the probability mass below zero for posterior predictions of $y(\bs,t)$ is consistently so small that this is nearly irrelevant in practice. \par
It is possible to include a regression on a set of covariates in both the binary and continuous model. This would not introduce a significant additional burden on inference for covariates of moderate dimension and could improve the predictive power of the model. The difficulty is obtaining covariates that are strong linear predictors and are spatially and temporally complete across the target domain. We did not include covariates in our analysis, but suggest some avenues for this in the Discussion.

\subsection{Covariance functions for forest inventory data}\label{sec:covfunc}

The choice of covariance function is critical for spatial-temporal modeling and can have drastic effects on predictions. The separable covariance functions of (\ref{eq:separblecov}) are simple and convenient, but we found them to be inadequate for the forest inventory data considered here and in other experimentation with FIA data. We hypothesize that separable functions perform poorly because fluctuations in forest attributes are the combined result of variation at different spatial scales. Much of the temporal flux in forest attributes is highly local, resulting from sources such as management decisions, soil, hydrology and small-scale disturbances. These are changes that cannot be confidently extrapolated over large spatial distances. Underlying these local variations are ecosystem-scale variations, originating from sources such as regional-scale climate, disturbance regimes, and species competition. With a single separable covariance function, likelihood-based inference pulls the unknown parameters to the dominant scale of variation (often the local scale), leaving other scales unrepresented. This is especially pertinent for FIA data, which contains repeat measurements at fixed locations. To maximize the likelihood with a separable covariance function, the model is often best off driving the spatial range to small values, fitting a nearly independent temporal Gaussian process to each fixed plot location, performing little spatial generalization. This is suboptimal from a prediction perspective, as it provides little power to spatially interpolate to unobserved locations. Further, a primary objective of forest inventories is inference on large area averages for which the neglected ecosystem-scale variation is highly relevant.\par
We propose to represent the Gaussian process as a sum of independent component Gaussian processes, representing variation at different scales. We assign a separable and exponential covariance function to each of the individual components,
\begin{align}
    w(\bs,t) &= \sum_{\ell=1}^L w_\ell(\bs,t),\\
    K(\bs,t) &= \sum_{\ell=1}^L \sigma_\ell^2\exp\left(-\frac{\ds}{\phi_\ell}\right)\exp\left(-\frac{\dt}{\lambda_\ell}\right), \label{eq:sumofseps}
\end{align}
giving $3L$ unknown parameters, $\btheta = [\sigma_1~\phi_1~\lambda_1~\cdots~\sigma_L~\phi_L~\lambda_L]^T$. The trio of parameters assigned to each individual component are exchangeable in $\pi(\bw|\btheta)$, so identifiability must be enforced through prior $\pi(\btheta)$ or some other form of regularization. We use $L = 2$ components with the local-scale vs.~ecosystem-scale interpretation discussed above, where informative yet defensible priors on spatial ranges, $\phi_1,\phi_2$, can distinguish the components.\par
\begin{figure}[htb]
    \centering
    \includegraphics[width=\linewidth]{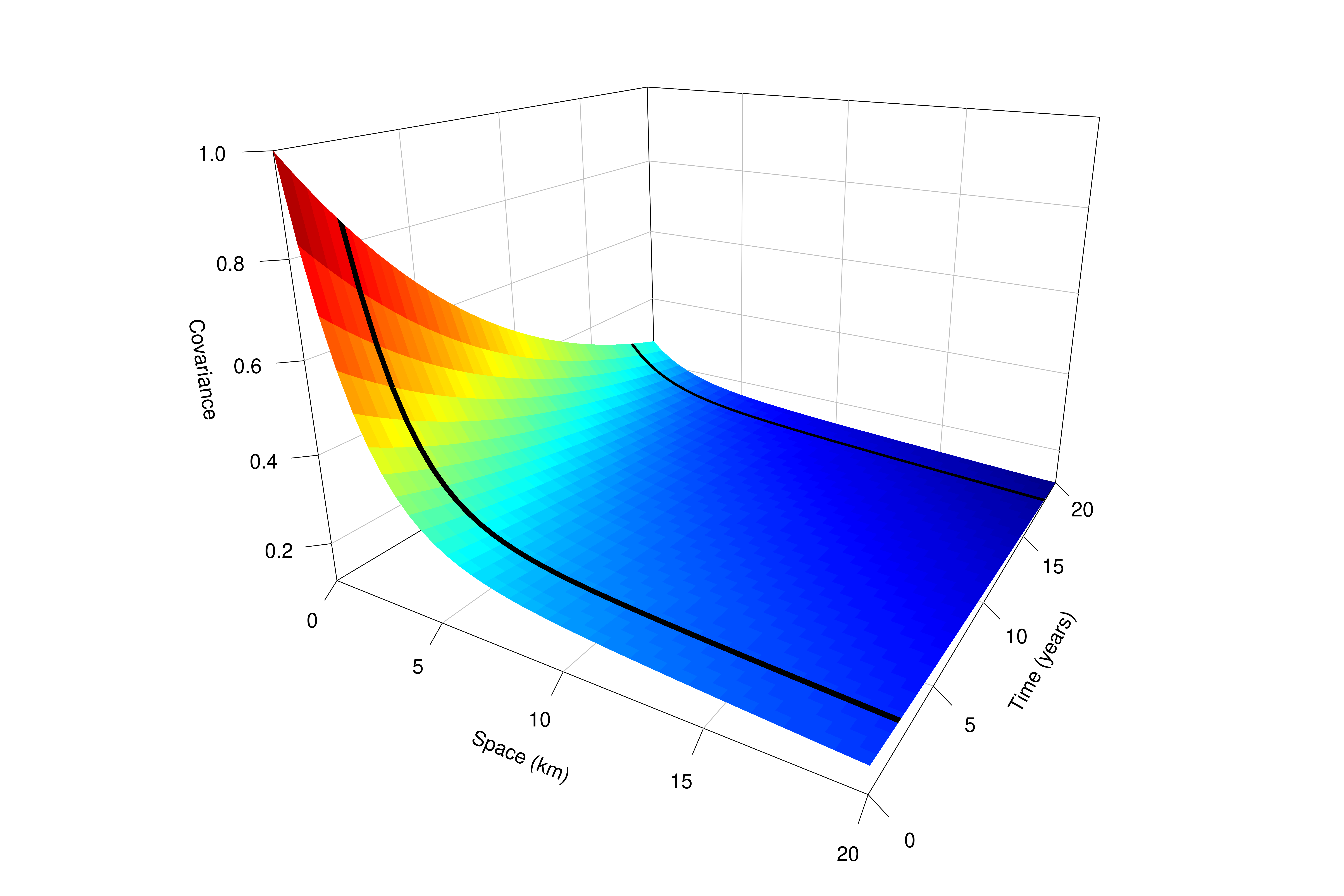}
    \caption{A sum of two separable covariance functions (eqn.~\ref{eq:sumofseps}, $L=2$) with $(\sigma_1, \phi_1, \lambda_1) = (\sqrt{0.25}, ~50~\text{km},~50~\text{years})$ representing a ecosystem-scale process and $(\sigma_2, \phi_2, \lambda_2) = (\sqrt{0.75}, ~2~\text{km},~10~\text{years})$ representing a local-scale process. The two black curves at time lags $\dt = 2,18~\text{years}$ differ beyond a multiplicative constant, making the resulting covariance function non-separable.}
    \label{fig:sumsepcorexample}
\end{figure}

Beyond the particular modeling application, a great degree of flexibility is gained with (\ref{eq:sumofseps}) over a single separable function. For $L > 1$ and distinct range parameters, the function is not separable in the sense that two fixed temporal lags will produce different linear combinations of the exponentials $\exp(-\ds/\phi_\ell),~\ell=1,\ldots,L$, which are not scalar multiples of each other (Figure \ref{fig:sumsepcorexample}).

\subsection{Inference and prediction}\label{sec:inference}
Let $\mathcal{O} \subset \calS \times \calT$ be the set of space-time points where $z(\bs,t)$ was observed, where $\bz_\mathcal{O}$ is the vector of observed binary values and $\bw_{z,\mathcal{O}}$ is the latent process at these points. Let $\mathcal{O}^*\subset\mathcal{O}$ be the forested subset where $y(\bs,t)$ was observed, with $\by_{\mathcal{O}^*}$ and $\bw_{y,\mathcal{O}^*}$ being corresponding observations and latent process vectors. We use $L=2$ components for $w_z(\bs,t)$ and $w_y(\bs,t)$, defining $\btheta_z$ and $\btheta_y$ as the respective sets of six covariance parameters. The posterior distributions of our unknown quantities, up to a normalizing constant, are
\begin{align}
    p(\bw_{z,\mathcal{O}}, \btheta_z, \alpha_z | \bz_\mathcal{O}) &\propto f(\bz_\mathcal{O}|\bw_{z,\mathcal{O}}, \alpha_z)\pi(\bw_{z,\mathcal{O}}|\btheta_z)\pi(\btheta_z)\pi(\alpha_z),\label{eq:zpost}\\
    p(\bw_{y,\mathcal{O}^*}, \btheta_y, \alpha_y, \tau | \by_{\mathcal{O}^*}) &\propto f(\by_{\mathcal{O}^*}|\bw_{y,\mathcal{O}^*}, \alpha_y, \tau^2)\pi(\bw_{y,\mathcal{O}^*}|\btheta_y)\pi(\btheta_y)\pi(\alpha_y)\pi(\tau),\label{eq:ypost}
\end{align}
which are structurally similar, except that the likelihood for $\bz_\mathcal{O}$ is Bernoulli and the likelihood for $\by_{\mathcal{O}^*}$ is normal with standard deviation $\tau$. We use independent Gamma priors for the strictly positive covariance parameters, $\btheta_z, \btheta_y$, and deviation parameter $\tau$. For the interpretability of the Gamma prior and ease of specification, we use a mean and standard deviation parameterization, where relative to the typical shape and rate parameterization we have
\begin{equation}
    \text{shape} = \frac{\text{mean}^2}{\text{sd}^2},\quad \text{rate} = \frac{\text{mean}}{\text{sd}^2}.
\end{equation}
For the covariance parameters with $L>1$, the priors should be informative enough to prevent interchangeable components. We enforce distinction primarily through the spatial ranges, ensuring prior probability $\Pr(\phi_1>\phi_2)\approx 1$, making the first component the larger scale process. Finally, we assign normal priors to the intercepts $\alpha_z$ and $\alpha_y$.\par
As is typical in Bayesian inference, predictive distributions and summaries of model unknowns require the evaluation of intractable integrals, requiring numerical methods. We use a Gibbs sampler to draw Markov chain Monte Carlo (MCMC) samples from the posterior distributions in (\ref{eq:zpost}, \ref{eq:ypost}). Appendix \ref{app:gibbs} gives details for this sampler. These samples of model unknowns are used to compute samples from posterior predictive distributions. For a set of prediction coordinates, $\mathcal{P}$, $M$ random samples are generated from $p(\by_\mathcal{P}|\by_\mathcal{O^*})$ and $p(\bz_\mathcal{P}|\bz_\mathcal{O})$, then generating predictive samples for AGBD:
\begin{equation}
    \boldsymbol{b}_\mathcal{P}^{(m)} = g\left[\by_\mathcal{P}^{(m)}\right]^{-1} \cdot \bz_\mathcal{P}^{(m)} \sim p(\boldsymbol{b}_\mathcal{P}|\by_\mathcal{O^*},\bz_\mathcal{O}) ;\quad m=1,\ldots,M,
\end{equation}
where the inverse transformation $g^{-1}(\cdot)$ and multiplication are applied element-wise.

\subsection{Model criterion and comparison}

We use a 10-fold cross-validation as a method for model evaluation and comparison of the $L=2$ and $L=1$ models. A typical cross-validation, leaving out a random sample of plot measurements for prediction, would be unsuitable because many of the test measurements would have companion measurements in the training set from the same plot, thus predicting an observed location but unobserved time. Therefore, we randomly split the plot locations into 10 folds, iteratively withholding each fold and all the associated measurements for prediction, until all measurements have an out-of-sample predictive posterior distribution.\par 
We assessed the performance of the predictive distributions through a number of metrics. Omitting the space-time indexing below for convenience (e.g., $b_i=b(\bs_i,t_i)$), we examined the accuracy of the posterior expected values of AGBD:
\begin{equation}
    \mathrm{MSE} = \frac{1}{n}\sum_{i=1}^n\left(E[b_i|\cdots] - b_i\right)^2.
\end{equation}
The magnitude of the above mean squared error should be interpreted relative to the variance of the quantity being predicted, so we computed a predictive R-squared metric
\begin{equation}
    R^2=1-\frac{n\cdot\mathrm{MSE}}{\sum_{i=1}^n\left(b_i - \bar{b}\right)^2}~,
\end{equation}
where $\bar{b}$ is the sample mean of the measured biomass. We measured the performance of the continuous, positive biomass predictions and the Bernoulli forest/non-forest predictions separately using mean log predictive densities
\begin{equation}
\begin{aligned}
    \mathrm{MLPD}_y &= \frac{1}{n^*}\sum_{i=1}^{n^*}\log\left[p(y_i|\cdots)\right],\\
    \mathrm{MLPD}_z &= \frac{1}{n}\sum_{i=1}^n\log\left[p(z_i|\cdots)\right],
\end{aligned}
\end{equation}
evaluating the respective predictive densities at the true test value, where $n^*$ is the number of forested, non-zero AGBD measurements. A larger $\mathrm{MLPD}$ indicates better performance, as the model is on average predicting higher posterior likelihoods at the true value.\par 

For the continuous observations, $y_i$, we also examined the coverage of the 95\% credible intervals, computed as the 2.5\% and 97.5\% quantiles of the predictive distribution. We did not extend this study to $z_i$ or $b_i$ because there is no way to define intervals of consistent probability for discrete or semi-continuous random variables.

Outside of the cross validation study, we computed a theoretical counterpart to the empirical variogram in (\ref{eq:empvario}),
\begin{equation}
    \gamma(B_k) = \frac{1}{|B_K|}\sum_{(\ds^{(ij)}, \dt^{(ij)})\in B_k} \left\{K(0,0)-K(\ds^{(ij)}, \dt^{(ij)})\right\},
\end{equation}
fixing the parameters of covariance function at their maximum a posteriori estimates, and comparing this to the empirical variogram to qualitatively assess the accuracy of the covariance function.


\section{Data analysis}\label{sec:analysis}
For the continuous, transformed process, $y(\bs,t)$, we used a cube-root transformation for the North Cascades and a square-root transformation for the Northeastern Highlands. All measurement times were converted to decimal years, e.g., January 15, 2002 is represented as $2002+15/365.25$. Our parameter priors are listed in Table \ref{tab:priors_y}  and \ref{tab:priors_z} of Appendix \ref{app:gibbs}. Broadly, we specified weakly informative priors for most parameters, except for the two spatial range parameters of each $L=2$ models. For all $L=2$ models, the ecosystem-scale spatial range, $\phi_1$, was assigned a Gamma prior with an expected value of 50 km and a standard deviation of 10 km. The local-scale component was assigned an expected value of 10 km and a standard deviation of 5 km. These two priors have little overlap in probability, which discourages the effective exchangability of the two components. For all models, after a burn-in period, we produced 20,000 MCMC samples from the posterior distribution (\ref{eq:zpost}, \ref{eq:ypost}).

\begin{table}[tp]
    \caption{Parameter posterior expected values and standard deviations for the two study areas.}
    \begin{subtable}{\textwidth}
    \centering
    \begin{tabular}{lcccc} \toprule
        & \multicolumn{2}{c}{$y(\bs,t)$} & \multicolumn{2}{c}{$z(\bs,t)$} \\
        \cmidrule(lr){2-3}\cmidrule{4-5}
         & $L = 2$ & $L = 1$ & $L = 2$ & $L = 1$ \\ \midrule
         $\sigma_1$ & 0.71 (0.07) & & 6.10 (0.83) &\\
         $\phi_1$ (km) & 55 (9) & & 58 (11) & \\
         $\lambda_1$ (years) & 154 (38) & & 107 (32) & \\ \addlinespace[5pt]
         $\sigma_2$ & 1.08 (0.02) & 1.35 (0.03) & 9.07 (0.68) & 13.02 (1.36)\\
         $\phi_2$ (km) & 0.43 (0.16) & 2.66 (0.17) & 2.96 (0.46) & 8.73 (0.87) \\
         $\lambda_2$ (years) & 374 (70) & 283 (41) & 295 (55) &  272 (50) \\ \addlinespace[5pt]
         $\tau$ & \num{9e-2} (\num{8e-3}) & \num{9e-2} (\num{9e-3}) & & \\ \bottomrule
    \end{tabular}
    \caption{North Cascades}
    \label{tab:cascadespartable}
    \end{subtable}
    
    \vspace{10pt}
    
    \begin{subtable}{\textwidth}
    \centering
    \begin{tabular}{lcccc} \toprule
        & \multicolumn{2}{c}{$y(\bs,t)$} & \multicolumn{2}{c}{$z(\bs,t)$} \\
        \cmidrule(lr){2-3}\cmidrule{4-5}
         & $L = 2$ & $L = 1$ & $L = 2$ & $L = 1$ \\ \midrule
         $\sigma_1$ & 1.01 (0.05) & & 12.65 (0.77) &\\
         $\phi_1$ (km) & 66 (8) & & 79 (10) & \\
         $\lambda_1$ (years) & 150 (19) & & 462 (61) & \\ \addlinespace[5pt]
         $\sigma_2$ & 2.13 (0.02) & 2.45 (0.02) & 14.92 (0.47) & 29.43 (1.03)\\
         $\phi_2$ (km) & 0.47 (0.07) & 1.98 (0.05) & 1.06 (0.17) & 11.95 (0.65) \\
         $\lambda_2$ (years) & 48 (1) & 58 (1) & 539 (64) &  919 (79) \\ \addlinespace[5pt]
         $\tau$ & \num{1e-3} (\num{1e-3}) & \num{3e-3} (\num{3e-3}) & & \\ \bottomrule
    \end{tabular}
    \caption{Northeastern Highlands}
    \label{tab:highlandspartable}
    \end{subtable}

    \label{tab:partable}
\end{table}

The posterior estimates of the parameters for the $L=2$ and $L=1$ models across both ecoregions are listed in Table \ref{tab:partable}. The $L=2$ model captures ecosystem-scale spatial variation with spatial ranges from 55 -- 79 km, as well as a more dominant, in terms of the associated standard deviation parameter, local-scale variation with spatial ranges from 0.5 -- 3 km. The $L=1$, having only one separable component, gravitates toward this local-scale variation, with a maximum spatial range of 2.7 km for the continuous $y(\bs,t)$ and 12 km for the Bernoulli $z(\bs,t)$. The sole spatial ranges inferred by the $y(\bs,t),~L=1$ models are short relative to the spatial density of FIA plots, leaving most possible prediction locations across the ecoregions with one or no plots within their inferred spatial ranges. All models for $\by(\bs,t)$ inferred $\tau$, the standard deviation of the random variation around the mean, to be effectively zero, inferring that all variation has some spatial and temporal coherence. The model has the power to confidently assert this due to the repeat measurements at identical plot locations.

\begin{figure}[htbp]
    \centering
    \includegraphics[width=\linewidth]{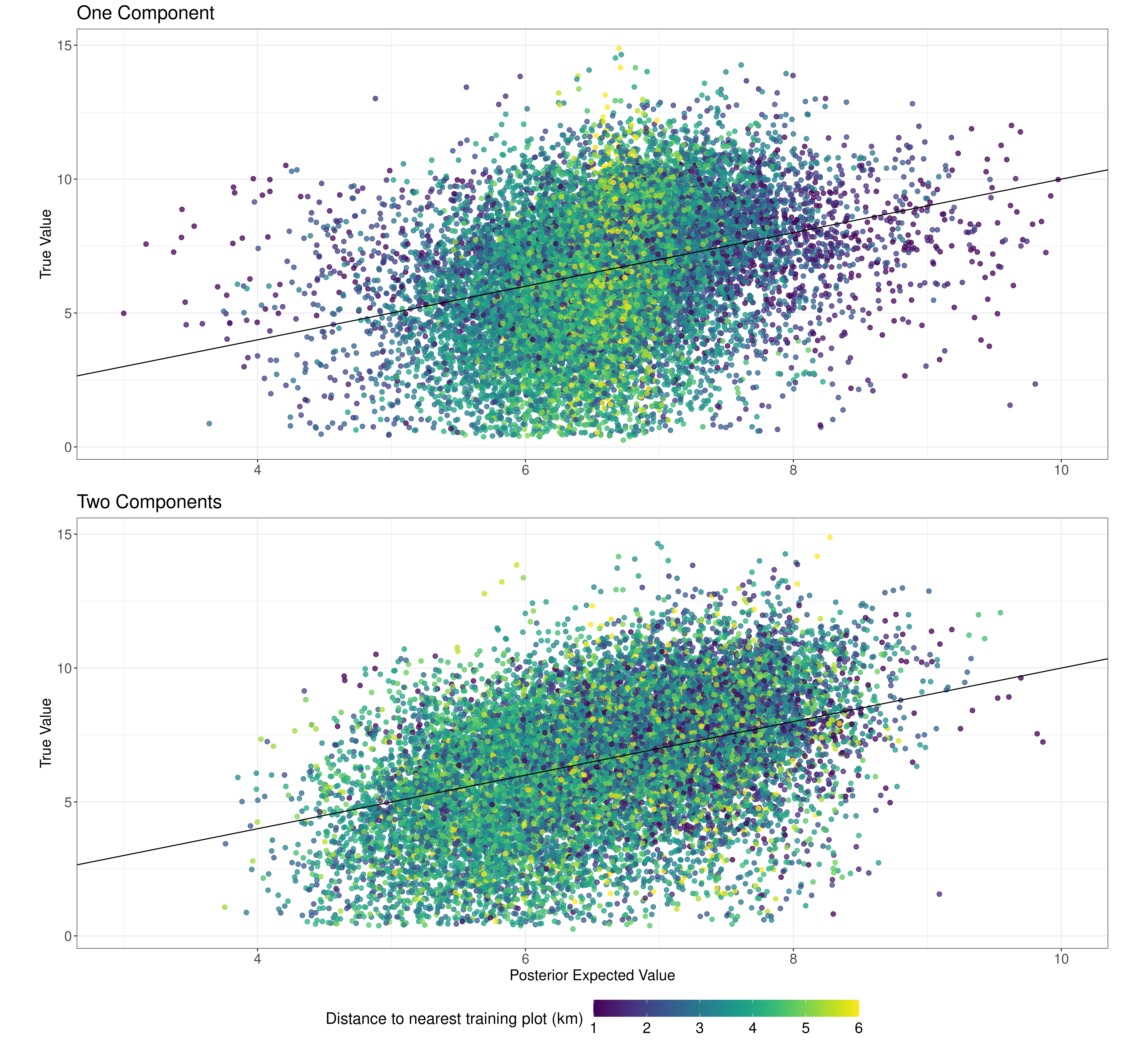}
    \caption{From the cross-validation for the Northeastern Highlands, posterior expected values of $y(\bs,t)$, the transformed AGBD in units $\sqrt{\text{Mg/ha}}$, versus the true values. The $L=1$ model depicts only spatially local variation, with predictions gravitating toward the intercept parameter for prediction locations distant from training plots. The $L=2$ model depicts broader spatial patterns, providing more informative predictions at all nearest-neighbor distances.}
    \label{fig:cvscatter}
\end{figure}

The cross-validation study showed superior predictive performance from the $L=2$ models. When comparing the posterior expected AGBD to the true AGBD for the $L=2$ and $L=1$ models, the predictive R-square values were 0.16 versus 0.25 for the North Cascades and 0.05 versus 0.14 for the Northeast Highlands. The $L=1$ model struggles to make informed predictions at any location far from an observed plot, often defaulting to predicting the intercept parameters $\alpha_y,~\alpha_z$ (Figure \ref{fig:cvscatter}).

The improved performance appears to be consistent across the $y(\bs,t)$ and $z(\bs,t)$ models. Comparing the $L=2$ and $L=1$ models, the $\mathrm{MLPD}$ for $y(\bs,t)$ was $-4.52$ and $-4.56$ respectively for the North Cascades and $-5.51$ and $-5.57$ for the Northeastern Highlands. The $\mathrm{MLPD}$ for the $z(\bs,t)$ model was $-0.36$ and $-0.38$ for the North Cascades and $-0.32$ and $-0.34$ for the Northeastern Highlands. 

Both the $L=2$ and $L=1$ models had appropriate coverage rates for the predictive 95\% credible intervals of $y(\bs,t$), with rates of $95.1\%$ and $93.7\%$ respectively for the North Cascades and $95.1\%$ and $94.9\%$ for the Northeastern Highlands.

\begin{figure}[p]
    \centering
    \includegraphics[width=0.86\linewidth]{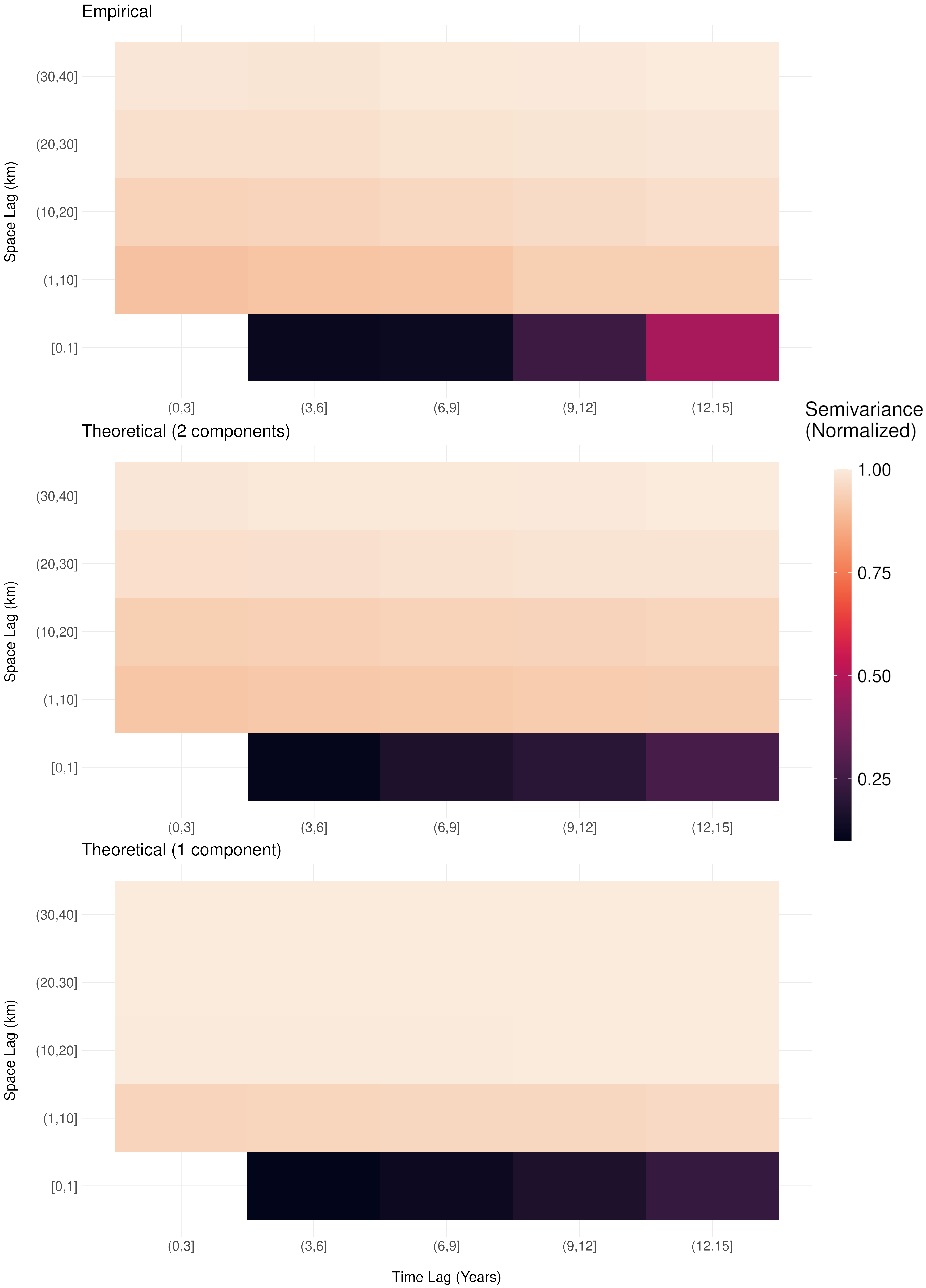}
    \caption{A comparison of the empirical semivariogram to theoretical semivariograms from the $L=2$ and $L=1$ models over the Northeastern Highlands. The $L=2$ model captures the sharp and then slow increase in variance across growing spatial lags, whereas the $L=1$ model captures only the sharp increase associated with local variation, quickly saturating to no covariance between pairs of locations.}
    \label{fig:variograms}
\end{figure}

We further illustrate the difference between the $L=2$ and $L=1$ models with the semivariograms (Figure \ref{fig:variograms}). We normalized each semivariogram by dividing by the maximum bin value to unify scales, as the standard deviation parameter for Mat\'ern class covariance functions is not individually identifiable \citep{zhang2004inconsistent}. The empirical semivariogram shows a sharp increase in variation due to local variation, followed by a slow increase over increasing spatial distances. The $L=2$ covariance function reflects this pattern, while the $L=1$ function captures only the local variation followed by immediate saturation.

\begin{figure}[ht]
    \centering
    \includegraphics[width=\linewidth]{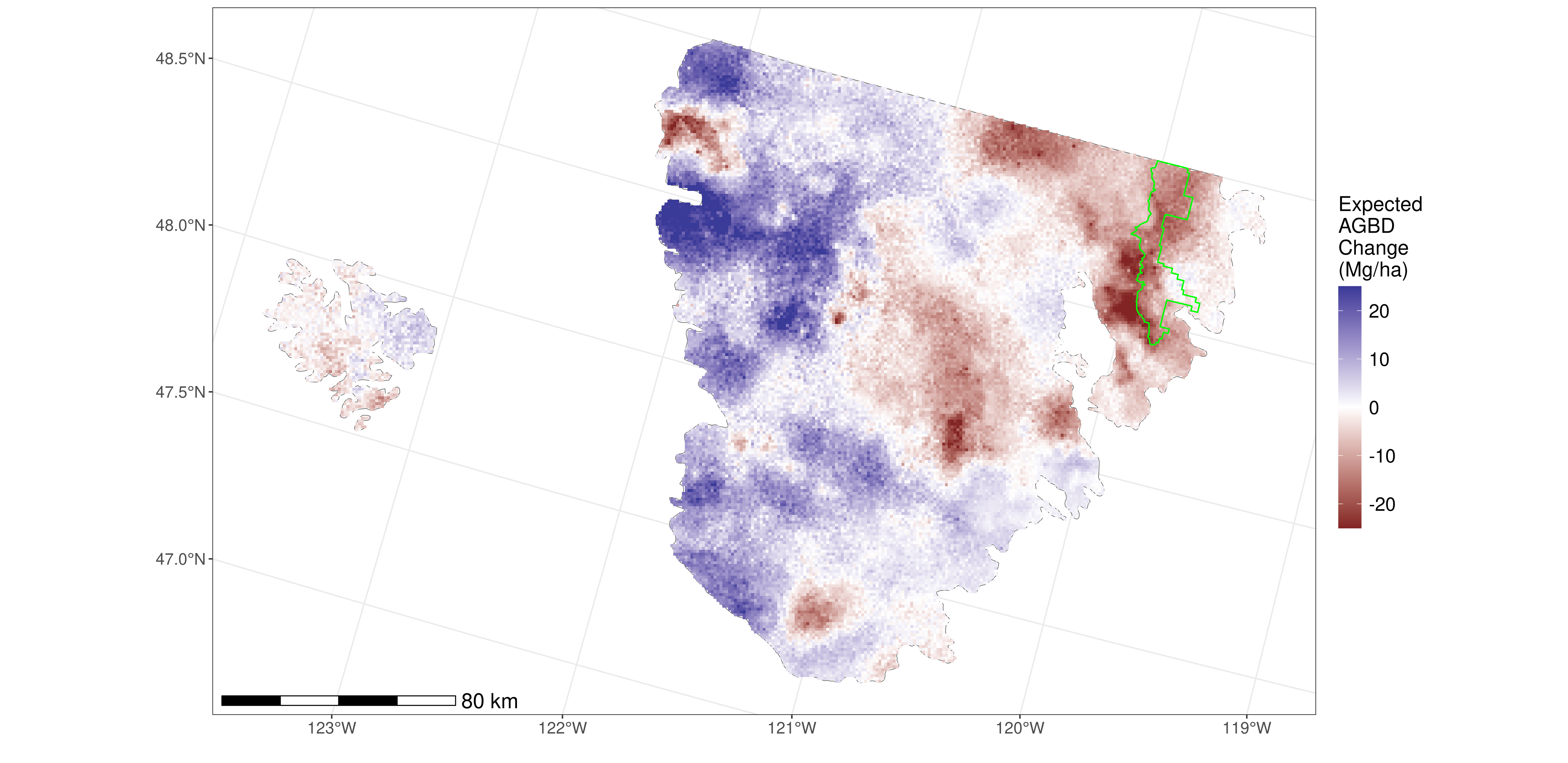}
    \caption{Posterior expected values of AGBD change from 2001 to 2021 (latter minus the earlier) sampled at a 1 km grid. Spatial-temporal models can produce predictions at the observation level on a spatial-temporal continuum, which can then be aggregated to an arbitrary target estimation area at arbitrary time points. Further, observation-level maps like the above provide information as to where within an estimation area the model is attributing gain or loss.}
    \label{fig:cascadeschange}
\end{figure}

\begin{figure}[htbp]
    \centering
    \includegraphics[width=\linewidth]{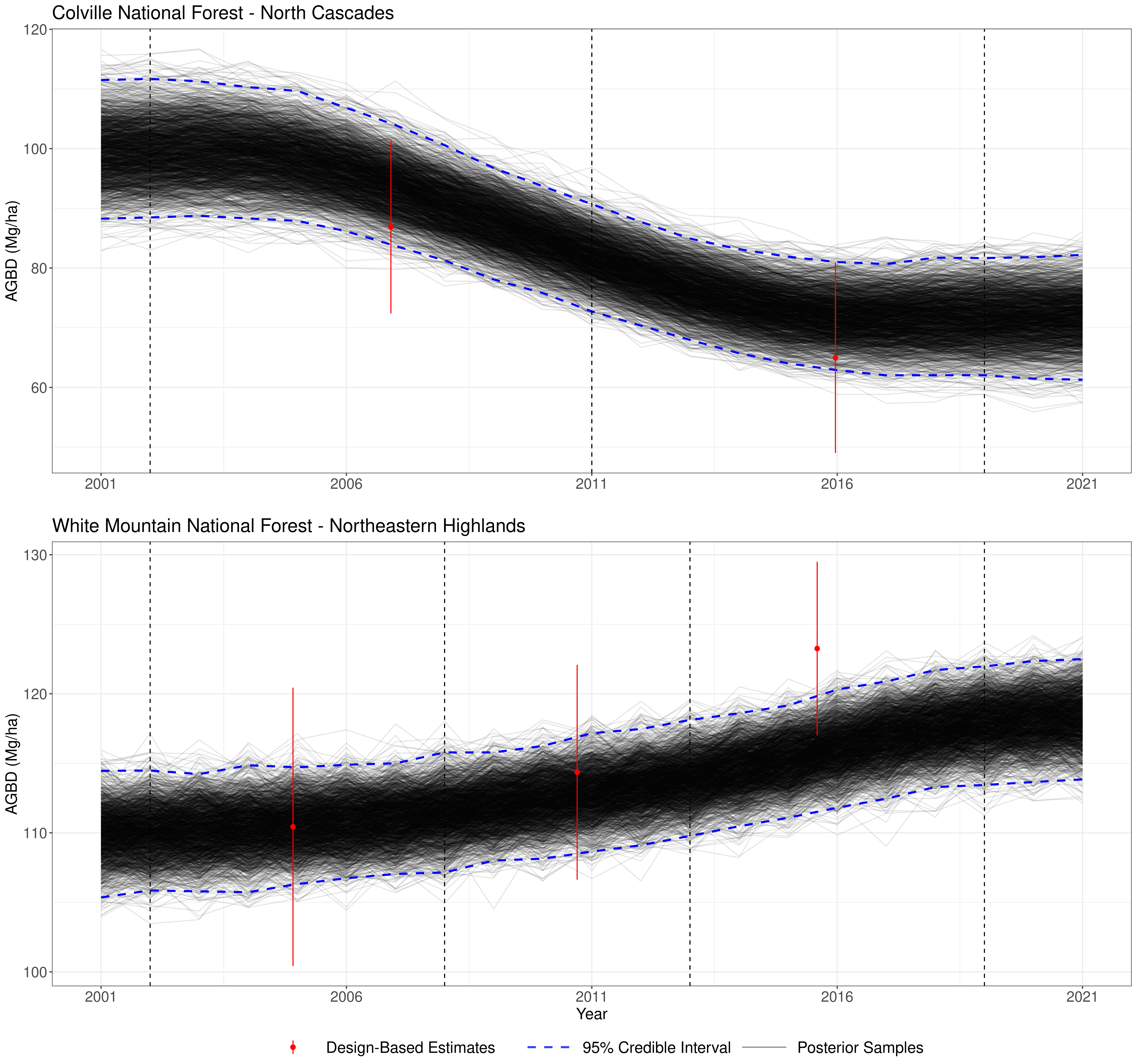}
    \caption{The posterior distribution, approximated with 2,000 posterior samples, for AGBD within our target estimation areas on a temporal continuum. Also pictured are design-based estimates using a Horvitz-Thompson estimator using complete cycles of FIA measurements (notice the $\sim\,$10 year spacing for Colville and $\sim\,$5 year spacing for White Mountain). The design-based estimates are centered horizontally at the average measurement year of the generating plots, with vertical ribbons spanning two estimate standard deviations above and below. The dashed vertical lines are at the first and last measurement years of the cycles.}
    \label{fig:nfstrends}
\end{figure}

Using the $L=2$ model, we produced a joint posterior predictive distribution for locations and times on a regular grid with a 1 km spatial density and an annual temporal density. For this stage, we thinned our MCMC sample set to 2,000 samples for memory efficiency. Figure \ref{fig:cascadeschange} shows this distribution summarized as the expected change from 2001 to 2021, $\E[b(\bs,2021)-b(\bs,2001)]$, for the North Cascades, depicting broad spatial patterns of both AGBD gain and loss. Samples from the joint posterior distribution can be collapsed to arbitrary spatial areas, which allows inference on average AGBD levels for a target area over time. Considering the National Forests we chose as our target areas (Figure \ref{fig:nfstrends}), it is estimated that the Colville National Forest in the North Cascades has sustained a marked decline in AGBD during the 20 year period, with an expected loss of 28 Mg/ha and a 95\% probability of a loss between 17 and 39 Mg/ha. This constitutes an expected decline of 28\% from the baseline levels of 2001. On the other hand, White Mountain National Forest in the Northeastern Highlands has an expected gain of 8 Mg/ha and a 95\% probability of a gain between 5 and 12 Mg/ha. We also used complete cycles within the considered time span to produce design-based estimates with a Horvitz-Thompson estimator \citep{horvitz1952generalization}, assuming simple random sampling. The design-based estimates broadly agree with the trends inferred by the spatial-temporal model, but are updated at approximately 5 or 10 year increments. The model-based approach can infer changes at subcycle time intervals: for example, Colville National Forest is expected to have lost 6 Mg/ha across 2008 -- 2010, with a 95\% probability of loss between 1 and 10 Mg/ha, and a 99\% probability that some loss occurred.

\section{Discussion}

Spatial-temporal models, and LGMs in particular, allow internally coherent estimates of forest population status and change at arbitrary spatial and temporal domains. This allows valid inference for a wide variety of objectives without being categorically limited by the sampling design. However, models do require validation in a way the design-based estimates do not. Design-based estimators are justified by the sampling design and purely mathematical arguments about the resulting properties of estimators. Models and model inference can be internally coherent given the assumptions, but must be empirically justified using the data, which is often done through cross-validation studies. A difficulty for validating models for forest inventory data is that the quantity we often seek to predict, area-averages of forest attributes, is not provided by the data. We can only test the performance of plot-level predictions and hope this reflects the performance of area-level predictions. For instance, the 95\% credible intervals for the LGM plot-level predictions of $y(\bs,t)$ had coverages of $\sim$95\%, but it is unclear that implies appropriate coverage for area-level credible intervals. Realistic simulation of artificial forest populations, possibly through generative machine learning methods, could be useful as sandboxes for model testing \citep[see, e.g., ][]{white2024kabaab}.

The primary goal of this work was to demonstrate rigorous inference on time-specific and change quantities, and not improved precision over the design-based estimates. But this does happen naturally, due to the borrowing of information from outside the spatial-temporal bounds of the target estimation area. In Figure \ref{fig:nfstrends}, the 95\% credible intervals of the model posterior distribution are considerably narrower than the 95\% confidence intervals of the design-based estimates, despite the model predictions representing time-specific quantities and the design-based estimates assuming temporal stasis across the cycle. A caveat is these metrics of uncertainty have different statistical interpretations (frequentist resampling from the sampling design vs.~Bayesian posterior beliefs), but we argue the practical interpretation by forest managers, policy makers, and other stakeholders will be the same. Further, the precision of the model predictions could be increased drastically by the inclusion of covariates. As mentioned previously, the primary obstacle is obtaining spatially and temporally complete covariates across a study domain. Harmonic coefficients derived from passive optical satellites such as the Landsat series seem especially promising to this end \citep{zhu2014continuous, wilson2018harmonic}. Computing these coefficients requires a considerable amount of technical expertise and computational power. For instance, the relevant Landsat data for our study areas and time domain is on the scale of terabytes. Free cloud computing platforms like Google Earth Engine provide the requisite computational tools, but the technical execution is still not trivial. This is a potentially fruitful area of future work.

For our demonstration, we settled on a covariance function composed of a sum of $L=2$ separable components, showing decidedly superior performance over a single separable function. We did not test models with $L>2$ components, partially due to the excellent match between the theoretical and empirical variograms with $L=2$ (Figure \ref{fig:variograms}), and partially due to the burden of repeatedly fitting and cross-validating models with increasing components. We tested other non-separable covariance functions, such as those from the Gneiting class \citep{gneiting2002nonseparable}, which have the advantage of depicting non-separability through a parameter to be inferred during model training, as opposed to comparing distinct and nested models with sum-of-separable covariance functions. However, we saw no differences in marginal likelihoods or posterior predictions when comparing a single Gneiting class function to a single separable function. As shown in the current study and in some prior investigations looking at spatial only data \citep[see, e.g.,][]{banerjee2007multires}, the processes used to approximate the latent forest inventory data generating process should accommodate multiple spatial and temporal scales. A formulation of this characteristic more compact than simply summing covariance functions would be useful.

Inference for LGMs can be complex and computationally demanding. We wrote a Gibbs sampler to draw random samples from the posterior distribution, using the NNGP to alleviate the demand of the Gaussian process. However, there are statistical software packages that can substantially reduce the technical programming required. For example, package `R-INLA' \citep{blangiardo2013spatial, rue2017bayesian} can fit the LGMs proposed in this work, using integrated nested laplace approximations for inference \citep{rue2009approximate} and an SPDE representation of the Gaussian process \citep{lindgren2011explicit}, with only a handful of lines of code. We would expect similar performance from an R-INLA implementation, and our decision to write the algorithms ``from scratch'' was based purely on a desire to increase familiarity with the tools.

\section{Conclusion}

We demonstrated latent Gaussian process models as a flexible tool to predict forest attributes and change thereof at arbitrary locations and times, addressing key limitations of design-based inference. Our results show that it is critical for the covariance function to account for variation at multiple scales, lest the analysis be dominated by short-scale variation that cannot be confidently interpolated. Future work includes leveraging current and historical satellite data to improve precision and enable confident predictions and finer spatial scales.

\section*{Data and Code}

FIA data is available for download at the FIA DataMart (\url{https://research.fs.usda.gov/products/dataandtools/tools/fia-datamart}) or through \texttt{R} package `FIESTA' \citep{Frescino_2023}. All algorithms used in this study are available as a \texttt{Julia} package at \url{https://github.com/PaulBMay/SpaceTimeMultiscale.jl}.

\section*{Funding} This work was supported by the USDA Forest Service [\#24-JV-11242305-111], National Science Foundation DEB-2213565 and DEB-1946007, and National Aeronautics and Space Administration CMS Hayes-2023. The findings and conclusions in this publication are those of the authors and should not be construed to represent any official US Department of Agriculture or US Government determination or policy.

\bibliographystyle{apalike}
\bibliography{references}

\appendix

\section{Nearest Neighbor Gaussian Process}\label{app:nngp}

Let $\mathcal{O} = \{(\bs_1,t_1),\ldots,(\bs_n,t_n)\}$ be the set of observed locations and times, and $\bw_\mathcal{O}$ be a vector representing the Gaussian process at these indices. Evaluating the multivariate normal prior, $\pi(\bw_\mathcal{O})$, and conditional distributions such as $p(\bw_\mathcal{P}|\bw_\mathcal{O})$, requires the solution of $n\times n$ linear systems involving $\bSigma$, where $[\bSigma]_{ij}=K\left((\bs_i,t_i),(\bs_j,t_j)\right)$, which will become prohibitive for large $n$. The NNGP \citep{datta2016hierarchical} is an extension of the Vecchia likelihood approximation \citep{vecchia1988estimation},
\begin{align}
    \pi(\bw_\mathcal{O}) &= \pi(w_1,\ldots,w_n)\nonumber\\
    &=\pi(w_1)\pi(w_2|w_1)\cdots\pi(w_n|w_1,\ldots,w_{n-1})\nonumber\\
    \approx \tilde{\pi}_{NN}(\bw_\mathcal{O}) &= \pi(w_1)\cdots\pi(w_i|\bw_{N(i)})\cdots\pi(w_n|\bw_{N(n)}), \label{eq:vecchia}
\end{align}
which truncates the conditioning for each $w_i$ at the subset of indices $\{(\bs_1,t_1)\ldots,(\bs_{i-1},t_{i-1})\}$ that are `nearest' to $(\bs_i,t_i)$, calling this set $N(i)$. The number of nearest neighbors is fixed at $|N(i)| = m,~\forall i\geq m$. The Vecchia approximation takes advantage of the screening property of many covariance functions, where the process is conditionally independent of more distant indices given the observation of nearer indices \citep{stein2002screening}. Instead of working with dense covariance matrix $\bSigma$, we compute the sparse precision matrix, $\bQ$, implied by (\ref{eq:vecchia}). Precision $\bQ$ will have at most $m$ non-zero entries per row, allowing us to use efficient sparse matrix routines for the requisite computations. The Vecchia approximation can be extended to a Gaussian process outside $\mathcal{O}$ by defining $\bw_\mathcal{P}=\bB_\mathcal{P}\bw_\mathcal{O}+\boldeta_\calP$, where $\bB_\mathcal{P}$ is a projection matrix dependent on $K(\cdot,\cdot)$, and $\boldeta_\calP\sim\mathrm{MVN}(\boldsymbol{0},\boldsymbol{F}_\calP)$, where $\boldsymbol{F}_\calP$ is a diagonal matrix, also dependent on $K(\cdot,\cdot)$. For more details on the structure and computation of $\bQ,\,\bB_\calP,\,\boldsymbol{F}_\calP$, see \cite{datta2016hierarchical} and \cite{finley2019efficient}.

For spatial-temporal data, the concept of a nearest neighbor is not as clear as in a purely spatial setting, where we could simply examine $\|\bs - \bs'\|$. We set $m=25$ and selected $N(i)$ and the $m$ nearest spatial neighbors over space, breaking ties by distance in time. We favored spatial distance in our neighbor selection, as we expected far more dynamism over space than time in forest variables. A more sophisticated, data-inferred scheme for neighbor selection for spatial-temporal data was studied in \cite{datta2016nonseparable}.

\section{Priors and Gibbs Samplers}\label{app:gibbs}

\subsection{Continuous process $y(\bs,t)$}\label{app:gibbs_y}

For convenience, let $\calO$ represent the set of observed locations and times for $y(\bs,t)$, though note we used $\calO^*$ in the body of the paper and that these indices will be a subset of the observation indices used in the next section for $z(\bs,t)$. We also drop all $y$ subscripts, though again note these parameters and effects are distinct from those in the next section. Let $\bQ(\btheta)$ be the prior precision matrix for $\bw_\calO$ with is dependent on the covariance parameters $\btheta$. The posterior distribution of our model unknowns is 
\begin{equation}
 p(\bw_\calO, \btheta, \alpha, \tau | \by_{\mathcal{O}}) \propto f(\by_{\mathcal{O}}|\bw_{\mathcal{O}}, \alpha, \tau)\pi(\bw_{\mathcal{O}}|\btheta)\pi(\btheta)\pi(\alpha)\pi(\tau).\label{eq:ypost_app}
\end{equation}

\begin{table}[t]
    \caption{Priors used for the continuous $y(\bs,t)$ model. The intercept was assigned a normal prior while strictly positive covariance parameters were assigned Gamma priors. We used these same priors for both analyzed ecoregions.}
    \centering
    \begin{subtable}{\textwidth}
    \centering
    \begin{tabular}{l c c c c c c c c}\toprule
         & Normal & \multicolumn{7}{c}{Gamma} \\
         \cmidrule(lr){2-2}\cmidrule(lr){3-9}
         & $\alpha$ & $\tau$ & $\sigma_1$ & $\phi_1$ (km) & $\lambda_1$ (years) & $\sigma_2$ & $\phi_2$ (km) & $\lambda_2$ (years)\\
         Mean & 5 & 1 &  2 & 50 & 100 & 4 & 10 & 100 \\
         Standard Deviation & 10 & 1 & 1.9 & 10 & 90 & 3.9 & 5 & 90 \\ \bottomrule
    \end{tabular}
    \caption{$L=2$ model.}
    \end{subtable}

    \vspace{10pt}

    \begin{subtable}{\textwidth}
    \centering
    \begin{tabular}{l c c c c c}\toprule
         & Normal & \multicolumn{4}{c}{Gamma} \\
         \cmidrule(lr){2-2}\cmidrule(lr){3-6}
         & $\alpha$ & $\tau$ & $\sigma_1$ & $\phi_1$ (km) & $\lambda_1$ (years)\\
         Mean & 5 & 1 &  3 & 25 & 100 \\
         Standard Deviation & 10 & 1 & 2.9 & 10 & 90 \\ \bottomrule
    \end{tabular}
    \caption{$L = 1$ model.}
    \end{subtable}
    \label{tab:priors_y}
\end{table}
Our priors for parameters $\alpha,\tau,\btheta$ are given in Table \ref{tab:priors_y}. We used a Gibbs sampler to draw samples from our posterior distribution. The sampler proceeds as
\begin{enumerate}
    \item \textbf{Sample from $p(\alpha,\bw_\calO|\by_\calO,\btheta,\tau)\propto f(\by_{\mathcal{O}}|\bw_{\mathcal{O}}, \alpha, \tau)\pi(\bw_{\mathcal{O}}|\btheta)\pi(\alpha)$.} Both $\alpha$ and $\bw_\calO$ have normal priors and are linear in (\ref{eq:ymu}), so we sample them jointly from their conditional posterior, which is multivariate normal. Let
    \begin{equation}
        \bm_\mu = \begin{bmatrix} m_\alpha \\ \bzero \end{bmatrix},\qquad \bQ_\mu = \begin{bmatrix} q_\alpha & \bzero \\ \bzero & \bQ(\btheta) \end{bmatrix}, 
    \end{equation}
    be the prior mean and precision for $[\alpha~\bw_\calO]^T$. Define the design matrix $\bD=[\bone~\Id]$. Then the posterior mean and precision are
    \begin{align}
        \tilde{\bm}_\mu &= \tilde{\bQ}_\mu^{-1}\left[\bQ_\mu\bm_\mu + \frac{1}{\tau^2}\bD^T\by_\calO \right],\\ 
        \tilde{\bQ}_\mu &= \bQ_\mu + \frac{1}{\tau^2}\bD^T\bD. 
    \end{align}
    The posterior precision matrix is sparse, so a Cholesky decomposition can be efficiently obtained through sparse matrix routines.
    \item \textbf{Sample from $p(\btheta,\tau|\by_\calO)\propto \pi(\btheta)\pi(\tau)\int f(\by_{\mathcal{O}}|\bw_{\mathcal{O}}, \alpha, \tau)\pi(\bw_{\mathcal{O}}|\btheta)\pi(\alpha)\,d\alpha\,d\bw_\calO$}. We sample $[\btheta~\tau]$ jointly and marginalize over the Gaussian effects to reduce autocorrelation in the Monte Carlo samples. This is feasible because
    \begin{equation}
        \int f(\by_{\mathcal{O}}|\bw_{\mathcal{O}}, \alpha, \tau)\pi(\bw_{\mathcal{O}}|\btheta)\pi(\alpha)\,d\alpha\,d\bw_\calO = f(\by | \btheta)
    \end{equation}
    is multivariate normal with mean $\bD\bm_\mu$ covariance matrix $\bD\bQ_\mu^{-1}\bD^T+\tau^2\Id$. Linear systems and log-determinants for this covariance matrix can be efficiently computed using the Woodbury Matrix Identity and determinant lemma. We sampled from $p(\btheta,\tau|\by_\calO)$ using a Metropolis-Hasting step with a log-Multivariate Normal proposal distribution.
\end{enumerate}

\subsection{Bernoulli process $z(\bs,t)$}

The posterior distribution of our model unknowns is 
\begin{equation}
 p(\bw_\calO, \btheta, \alpha | \bz_{\mathcal{O}}) \propto f(\bz_{\mathcal{O}}|\bw_{\mathcal{O}}, \alpha)\pi(\bw_{\mathcal{O}}|\btheta)\pi(\btheta)\pi(\alpha),\label{eq:zpost_app}
\end{equation}
Our priors for parameters $\alpha,\btheta$ are given in Table \ref{tab:priors_z}.
\begin{table}[t]
    \caption{Priors used for the Bernoulli $z(\bs,t)$ model. The intercept was assigned a normal prior while strictly positive covariance parameters were assigned Gamma priors. The same priors were used for both analyzed ecoregions, except for the prior mean of $\alpha$, assigning 1.5 to the North Cascades and 2 to the Northeastern Highlands, reflecting our higher prior expected log-proportion of forest for the latter region.}
    \centering
    \begin{subtable}{\textwidth}
    \centering
    \begin{tabular}{l c c c c c c c }\toprule
         & Normal & \multicolumn{6}{c}{Gamma} \\
         \cmidrule(lr){2-2}\cmidrule(lr){3-8}
         & $\alpha$ & $\sigma_1$ & $\phi_1$ (km) & $\lambda_1$ (years) & $\sigma_2$ & $\phi_2$ (km) & $\lambda_2$ (years)\\
         Mean & 1.5, 2 & 4 & 50 & 100 & 8 & 10 & 100 \\
         Standard Deviation & $1/\sqrt{10}$ & 3.9 & 10 & 90 & 7.9 & 5 & 90 \\ \bottomrule
    \end{tabular}
    \caption{$L=2$ model.}
    \end{subtable}

    \vspace{10pt}

    \begin{subtable}{\textwidth}
    \centering
    \begin{tabular}{l c c c c}\toprule
         & Normal & \multicolumn{3}{c}{Gamma} \\
         \cmidrule(lr){2-2}\cmidrule(lr){3-5}
         & $\alpha$ & $\sigma_1$ & $\phi_1$ (km) & $\lambda_1$ (years)\\
         Mean & 1.5, 2 & 9 & 25 & 100 \\
         Standard Deviation & $1/\sqrt{10}$ & 8.9 & 10 & 90 \\ \bottomrule
    \end{tabular}
    \caption{$L = 1$ model.}
    \end{subtable}
    \label{tab:priors_z}
\end{table}
Our algorithm for the Bernoulli process must differ from the previous algorithm for the normal data model because conditional posterior $p(\alpha,\bw_\calO|\bz_\calO,\btheta)$ is not multivariate normal, and marginalizing over $\bw_\calO$ is no longer feasible. We use Poly\'a-Gamma data augmentation \citep{polson2013bayesian}, which introduces a vector of Poly\'a-Gamma variables, $\bomega$, to make the conditional posterior for $\alpha,\bw_\calO$ more tractable:
\begin{align}
    p(\alpha,\bw_\calO|\bz_\calO,\btheta) &\propto \int p(\alpha,\bw_\calO|\bz_\calO,\btheta,\bomega)\pi(\bomega)~d\bomega \\
    \text{where}\quad \alpha,\bw_\calO|\bz_\calO,\btheta,\bomega &\sim \mathrm{MVN}.
\end{align}
The Gibbs sampler proceeds as follows. 
\begin{enumerate}
    \item \textbf{Sample from $p(\alpha,\bw_\calO|\bz_\calO,\btheta,\bomega) \propto f(\bz_\calO|\bw_\calO,\alpha,\bomega)\pi(\bw_\calO|\btheta)\pi(\alpha)\pi(\bomega)$}. Define prior mean and precision, $\bm_\mu,\bQ_\mu$, and design matrix, $\bD$, the same as in Step 1 of Section \ref{app:gibbs_y}. Further, define the diagonal matrix $\bOmega = \mathrm{diag}\{\bomega\}$. Then $\alpha,\bw|\cdots$ is multivariate normal with mean and precision
     \begin{align}
        \tilde{\bm}_\mu &= \tilde{\bQ}_\mu^{-1}\left[\bQ_\mu\bm_\mu + \bD^T(\by_\calO - \boldsymbol{\frac{1}{2}}) \right],\\ 
        \tilde{\bQ}_\mu &= \bQ_\mu + \bD^T\bOmega\bD. 
    \end{align}
    \item \textbf{Sample from $p(\bomega|\alpha,\bw_\calO)$}, where $\omega_i|\alpha,\bw_\calO \sim \mathrm{PG}(1,\alpha+w_i)$ for $i=1,\ldots,n$. See \cite{polson2013bayesian} for a proof of this, as well as efficient algorithms for sampling Poly\'a-Gamma variables.
    \item \textbf{Sample from $p(\btheta|\bw_\calO)\propto\pi(\bw_\calO|\btheta)\pi(\btheta)$.} As mentioned before, marginalizing out $\bw_\calO$ is not feasible with a Bernoulli likelihood, so we must condition on the Gaussian effects. We sampled from $p(\btheta|\bw_\calO)$ using a Metropolis-Hastings step with a log-Multivariate Normal proposal distribution.
\end{enumerate}

\end{document}